\documentclass[aps,pra, twocolumn,nofootinbib,superscript address]{revtex4-2}
\usepackage[utf8]{inputenc}
\usepackage[english]{babel}
\usepackage[T1]{fontenc}
\usepackage{amsfonts}
\usepackage{amsmath}
\usepackage{physics}
\usepackage{hyperref}
\usepackage{amsthm}

\usepackage{tikz}
\usepackage{natbib}

\newcommand{\id}{\mathbb{I}}
\newcommand{\wk}{\overline{W}}
\newcommand{\wtilde}{\widetilde{W}}
\newcommand{\Q}{\overline{Q}}

\newcommand{\myblue}{black} 

\begin{document}

\title{\textcolor{\myblue}{Thermodynamically Optimal Protocols for Dual-Purpose Qubit Operations}}
\author{Joe Dunlop} \email[he/him/his ]{j.dunlop@exeter.ac.uk} \affiliation{Physics and Astronomy, University of Exeter, Exeter EX4 4QL, United Kingdom}

\author{Federico Cerisola}
\affiliation{Physics and Astronomy, University of Exeter, Exeter EX4 4QL, United Kingdom}
\affiliation{Department of Engineering Science, University of Oxford, Parks Road, Oxford, OX1 3PJ, United Kingdom}

\author{Jorge Tabanera-Bravo}\affiliation{Dept.~Estructura de la Materia, F\'isica T\'ermica y Electr\'onica and GISC, Universidad Complutense de Madrid. 28040 Madrid, Spain}

\author{Janet Anders} \affiliation{Physics and Astronomy, University of Exeter, Exeter EX4 4QL, United Kingdom} \affiliation{Institut für Physik, Potsdam University, 14476 Potsdam, Germany}

\date{\today}

\begin{abstract}
Information processing, quantum or classical, relies on channels transforming multiple input states to different corresponding outputs. Previous research has established bounds on the thermodynamic resources required for such operations, but no protocols have been specified for their optimal implementation. For the insightful case of qubits, we here develop explicit protocols to transform \textcolor{\myblue}{two} states in an energetically optimal manner. We first prove conditions on the feasibility of carrying out such transformations at all, and then quantify the achievable work extraction. Our results uncover a fundamental incompatibility between the thermodynamic ideal of slow, quasistatic processes and the information-theoretic requirement to preserve distinguishablity between different possible output states.
\end{abstract}
\maketitle

\section{Introduction}\label{intro}

\begin{figure}[b!]
    \centering
    \includegraphics[width=\columnwidth]{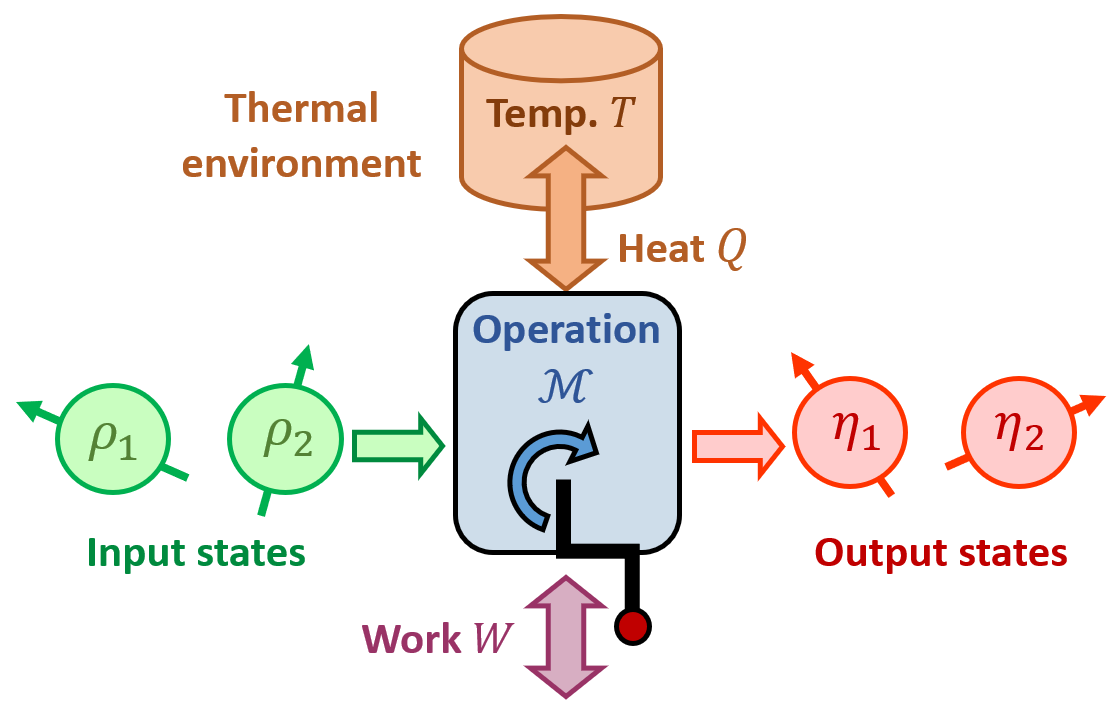}
    \caption{A \textcolor{\myblue}{dual-purpose qubit} operation transforms \textcolor{\myblue}{an input state, either $\rho_1$ or $\rho_2$ (occurring with probabilities $p_1, p_2$),} into the respective output \textcolor{\myblue}{$\eta_1$ or $\eta_2$}. This is implemented via some cyclic variation of the Hamiltonian, in a procedure which does not require knowledge of which of the states is present. During the process, the system exchanges heat with a fixed-temperature environment, and work with the driving field. Such operations are essential for information processing, and we would like to determine the energetic requirements.}
    \label{cartoon}
\end{figure}

In introducing entropy as an information measure, Shannon stressed an essential property of communication channels: that the message to be transmitted must belong to a set of possible messages, and that "the system must be designed to operate for each possible selection, not just the one which will actually be chosen since this is unknown at the time of design" \cite{shannon}. In general, information processing involves \textit{multipurpose operations} mapping each of a collection of possible input signals to their corresponding outputs via some standard protocol (Fig. \ref{cartoon}) \cite{Holevo_1998}. Signals must be encoded in the states of a physical system, with stored information manifested as entropy of the medium; and all operations on them must obey thermodynamic laws -- in particular the Clausius inequality, applied by Landauer to lower-bound the energetic cost of erasure in the presence of a thermal environment \cite{landauer,bennett03,zurek2003,Koji}.

Quantum information theory and the development of quantum technologies \cite{Deutsch} have driven a surge of interest in basic limits on thermodynamic performance \cite{Goold,Parrondo2015,Linden}, and the role of novel quantum resources such as coherence \cite{Anders,Francica,Smith,Camati_2019,Lostaglio2015,Korzekwa_2016,Scully} and entanglement \cite{delrio,bresque_2021,Hovhannisyan_2013,Huang_2013,Hewgill,williamson_2023,Perarnau_2015}. Beyond foundational interest, energy consumption and heat generation may be an important factor in the viability of future generations of hardware \cite{Auffeves,Gammaitoni_2015}.

The energetic requirements for transforming a given quantum state to another have been studied in some detail. If the entropy of the output state differs from the input, the transformation must involve interaction with a heat bath. Nonetheless, in the absence of additional constraints (e.g. imperfect control or finite duration \cite{Zhen,Mohammady,taranto}), there is no fundamental barrier to these processes being carried out reversibly, provided the protocol is optimised for the particular input state present.
Indeed, concrete protocols have been developed which in theory saturate the reversible limit, where the work extraction is equal to the reduction of system's free energy \cite{Takara,Anders_Giovannetti,Anders}.

On the other hand, relatively little attention has been given to the thermodynamics of multipurpose quantum operations, which map each of a collection of possible input states $\rho_n$ to respective outputs $\eta_n$; \textcolor{\myblue}{the prototypical example being a single-qubit operation with two input states (Fig. \ref{cartoon}).} While the set of inputs is fixed, it is not known \emph{which} of the inputs is present in a given instance \cite{Holevo_1998}. The task is now to find a multipurpose operation $\mathcal{M}$ that maps \emph{every} possible input state to its target output state. Constructing a one-size-fits-all protocol presents a more constrained problem: the optimal strategy for transforming $\rho_1$ to $\eta_1$ might fail to transform $\rho_2$ to $\eta_2$.

It has been shown by Bedingham and Maroney that in general, multipurpose operations necessarily incur an irreversible energy cost which is dissipated to the thermal environment \cite{maroney}, in addition to the reversible work implied by the Landauer bound. Chiribella et al have explored the nonequilibrium cost of carrying out multipurpose operations \textit{approximately}, using a resource-theoretic approach to lower-bound the number of pure qubit states consumed for a given accuracy of outputs \cite{Fei}.

However, key questions remain open: there is no prescription for \textit{how} \textcolor{\myblue}{an operation with two or more inputs} might be carried out in a thermodynamically optimal way. Moreover, no upper bound has been placed on the energy cost.

The present paper undertakes a detailed thermodynamic analysis of single-qubit operations with two possible input states and corresponding outputs: \textcolor{\myblue}{\textit{dual-purpose qubit operations}}. We introduce a family of explicit protocols for carrying out such operations, as well as means to optimise for thermodynamic performance. These results are applied in a numerical case study of work extraction from coherence, highlighting the scale of the work penalty, as well as qualitative differences from the single-input case. Finally, we derive an analytic bound on work extraction, which is complementary to Bedingham and Maroney's result, despite arising from an entirely different theoretical treatment.

Our bound on work extraction is related to an effective speed limit: in order to preserve the output states' mutual distinguishability, any thermal contact with the environment must be limited in duration, avoiding complete thermalisation. This is fundamentally at odds with the requirements for thermodynamic reversibility, where state transformations involving contact with the environment must happen extremely slowly, with the state remaining in thermal equilibrium. Our results call attention to the scale of the resulting energetic toll, and clearly illustrate the general principles preventing reversibility for multipurpose operations acting on systems of any size, with any number of inputs.

\section{Setup}\label{setup}

We will primarily consider the following scenario. We are given a qubit which has been prepared in either state $\rho_1$ or $\rho_2$, and tasked with transforming it to a corresponding output state $\eta_1$ or $\eta_2$, depending on the input. We know what the possible states are, and the respective probabilities of their occurrence, $p_1$ and $p_2$, but not which of the states is present in a given instance. The objective is to carry out the mapping exactly for both inputs via some shared protocol, whilst maximising the work extracted (or minimising work expended) in the process. If our method extracts $W_1$ of work in transforming $\rho_1$ to $\eta_1$, and $W_2$ in transforming $\rho_2$ to $\eta_2$, then we aim to maximise the mean work $\wk = p_1 W_1 + p_2 W_2$.

Crucially, from an information perspective it is not sufficient to map the average input to the average output. For example, consider a scenario where the input states are $\rho_1 = \ketbra{0}{0},\,\rho_2=\ketbra{1}{1}$ and each occur with probability $\frac{1}{2}$, so that the average input is $\overline{\rho} = \frac{\id}{2}$. If we consider only their action on the average state, then the qubit gates $\id,X,Y$ and $Z$ have an identical effect, all producing the output $\overline{\eta} = \frac{\id}{2}$. But clearly it would not do to replace $X$ with $\id$ in a circuit: the action on individual input states is important.

It will be assumed that we can vary the qubit Hamiltonian freely, and thereby control the system's unitary dynamics - subject to the condition that the Hamiltonian is reset to its initial value $H_0$ at the end of the protocol, so that the operation might be performed in a cycle (for example, to process a string of inputs). It is also assumed that if the Hamiltonian is held fixed at $H$, the qubit will thermalise over some finite timescale to the Gibbs state $\tau = \frac{1}{Z} \exp(-\frac{H}{k_\mathrm{B} T})$, where the temperature $T$ of the thermal environment is taken to be fixed throughout, and where $Z=\tr[\exp(-\frac{H}{k_\mathrm{B} T})]$ denotes the partition function.

As is standard practice in quantum thermodynamics \cite{Alicki,Goold,Anders_Giovannetti}, we define the work done by the system as equal to the reduction of its internal energy as a direct result of changing the Hamiltonian; and heat absorbed by the system as an increase in its internal energy due to a changing state:
\begin{equation}\label{workdefs}
    \begin{split}
        \dot{W} &= -\tr[\rho \dot{H}],\\
        \dot{Q} &= \tr[H \dot{\rho}].
    \end{split}
\end{equation}
We adopt the convention that positive work is done \emph{by} the system against the driving field, and positive heat is absorbed by the system from the thermal environment. Storing or otherwise utilising the extracted work is a separate issue \cite{Skrzypczyk2014,Perarnau_2015} not treated here.

The Clausius inequality dictates that the work yield of a thermodynamic process is no greater than the reduction of free energy $F(\rho) = \tr[H\rho] - k_\mathrm{B} T \,S(\rho)$ between the initial and final state (here $S(\rho) = -\tr[\rho\ln \rho]$ denotes the von Neumann entropy) \cite{Esposito_2011}. Applying this to our \textcolor{\myblue}{dual-purpose} qubit operation, the mean work extraction is bounded by the free energy difference between the average input state $\overline{\rho} = p_1 \rho_1 + p_2 \rho_2$ and the average output $\overline{\eta} = p_1 \eta_1 + p_2 \eta_2$:
\begin{equation}\label{clausius}
    \begin{split}
        \wk &\leq F(\overline{\rho}) - F(\overline{\eta})\\
        &= \tr[H_0(\overline{\rho} - \overline{\eta})] -k_\mathrm{B} T \left[S(\overline{\rho}) - S(\overline{\eta})\right].
    \end{split}
\end{equation}
We will refer to protocols which saturate \eqref{clausius} as thermodynamically reversible - in that the average output could be transformed back to the average input at zero net work cost. Note that the process might still be \emph{logically} irreversible: it might be impossible to deterministically recover $\rho_1$ from $\eta_1$ or $\rho_2$ from $\eta_2$ \cite{Sagawa_2014}. For example, Landauer erasure is logically irreversible but may saturate \eqref{clausius}.

The above already implies a thermodynamic compromise. If we were able to choose our protocol based on prior knowledge of which of the states $\rho_{1,2}$ was present, we could extract work equal to $F(\rho_{1,2}) - F(\eta_{1,2})$. However, due to the concavity of the von Neumann entropy, the ideal \textcolor{\myblue}{dual-purpose} work extraction given by \eqref{clausius} is generally less than the weighted average: $F(\overline{\rho}) - F(\overline{\eta}) \leq p_1\left(F(\rho_{1}) - F(\eta_{1})\right) + p_2\left(F(\rho_{2}) - F(\eta_{2})\right)$.

This is \emph{not} the compromise we will be exploring in detail here. Instead we will be investigating ways that a \textcolor{\myblue}{dual-purpose} protocol might approach the bound given by \eqref{clausius}, and analysing the reasons why even that bound cannot in general be saturated. We base our approach on a well-studied type of reversible protocol, which maps a single, known initial state to some other final state \cite{Anders}. Taking this as a starting point, we identify  necessary extensions for a \textcolor{\myblue}{dual-purpose} operation, and see where those extensions violate the conditions for thermodynamic reversibility.

\subsection{How are single-input operations carried out reversibly?}

It is instructive to consider how an operation with a single input $\rho$ and output $\eta$ can be implemented reversibly. A protocol which appears in various forms in the literature \cite{Takara,maroney,Smith,Anders} proceeds along the following lines:
\begin{enumerate}
    \item Begin with a system in the state $\rho$, with initial Hamiltonian $H_0$. Abruptly quench the Hamiltonian to $H_\rho = -k_\mathrm{B} T (\ln \rho + \ln Z\, \id)$. The system, still in state $\rho$, is in thermal equilibrium\footnote{\textcolor{\myblue}{This approach fails to work exactly if $\rho$ or $\eta$ are pure states, since $H_{\rho/\eta}$ would require an infinite energy gap. However, the operation can still be carried out \textit{approximately}, mapping between mixed states $\varepsilon$-close to $\rho$ and $\eta$, for a finite energy cost.}} at temperature $T$ with respect to $H_\rho$, since $\rho = \frac{1}{Z}\exp(-\frac{H_\rho}{k_\mathrm{B} T})$.
    \item Slowly and smoothly adjust the Hamiltonian from $H_\rho$ to $H_\eta = -k_\mathrm{B} T (\ln \eta + \ln Z\, \id)$, over a period much longer than the thermalisation timescale. The system undergoes quasistatic, isothermal evolution, tracking the instantaneous thermal state from $\rho$ to $\eta$.
    \item Quench the Hamiltonian back to $H_0$, quickly enough that the system undergoes negligible evolution and remains in the intended output state $\eta$.
\end{enumerate}
When the thermodynamic analysis is performed using Eq. \eqref{workdefs}, this method is found to saturate the Clausius inequality \eqref{clausius} for both classical and quantum states \cite{Takara,Anders}. Key to its reversibility is the fact that the processes are either adiabatic, involving no exchange of heat with the environment (steps 1 and 3); or quasistatic, with the state at all times in thermal equilibrium with the environment (step 2). The system never undergoes irreversible thermalisation from an initial state which is appreciably out of equilibrium \cite{Mohammady}.

Why can't the same approach be used for a multipurpose operation? Notice that the Hamiltonians $H_\rho$ and $H_\eta$ are fine-tuned to the input and output state. This can't be done simultaneously for more than one possible input, since if $\rho_1 \neq \rho_2$, then $H_{\rho_1}\neq H_{\rho_2}$.

Suppose that we optimise this method for the \emph{average} input and output, replacing $H_\rho$ and $H_\eta$ with $H_{\overline{\rho}}$ and $H_{\overline{\eta}}$ in the above. The system, regardless of whether it began in state $\rho_1$ or $\rho_2$, thermalises to $\overline{\rho}$ at the start of step 2, before slowly evolving from there to $\overline{\eta}$, finally remaining in that state as the Hamiltonian is reset in step 3.

This is unacceptable: we obtain the same output state $\overline{\eta}$ irrespective of the input. It does not suffice to produce the right output state only on average\footnote{Unless the objective is to reset the system to a standard state, for example $\eta_1 = \eta_2 = \ketbra{0}{0}$, as in Landauer erasure. In that case the above indeed represents an optimal, thermodynamically reversible strategy.}. A similar problem arises no matter what state we optimise for. In fact, any initial-state dependence is lost as soon as the system is allowed to completely thermalise, since the Gibbs state is uniquely determined by the Hamiltonian for a given temperature. This rules out any protocol involving quasistatic isothermal evolution, wherever we require \textcolor{\myblue}{two or more} distinct outputs. For multipurpose operations, we need a framework to treat nonequilibrium processes.

\subsection{Discrete quantum processes}

Our approach is to decompose the operation as a sequence of discrete steps which are either unitary (and involve only work) or thermalising (involving only heat transfer). This framework, introduced in \cite{Anders_Giovannetti}, provides a means to quantify heat and work for processes which start or end in nonequilibrium configurations, and recovers continuous trajectories in the limit of many steps, including reversible quasistatic ones. However, it is necessary to extend that framework to account for \textit{incomplete} thermalisation, for the reason outlined above. We take the primitive operations to be the following:

\textcolor{\myblue}{\paragraph{Unitaries combined with Quenches. } These encompass operations which involve zero heat transfer, which can be further decomposed into unitary evolution generated by a fixed Hamiltonian, $U = e^{-\frac{i}{\hbar} H t}$; and instantaneous quenches, which alter the Hamiltonian $H\mapsto H'$ without any immediate change in the qubit state. The action of the combined unitary and quench $\mathcal{U}$ is to map the state $\rho\mapsto U \rho U^\dagger$, while extracting work equal to $W = \tr[H\rho] - \tr[H' U \rho U^\dagger]$.}

\textcolor{\myblue}{\paragraph{Partial Thermalisations. } Conversely, these represent processes where zero work is done. A partial thermalisation $\mathcal{T}{:}\, \rho \mapsto \lambda \rho + (1{-}\lambda)\tau$ mixes the qubit state with the Gibbs thermal state, $\tau = \frac{1}{Z} \exp(-\frac{H}{k_\mathrm{B} T})$ for a given Hamiltonian $H$, with mixing parameter $0\leq\lambda\leq 1$. This represents a linear interpolation of the dynamics taking $\rho$ to $\tau$, and we need not consider any details of the environment other than its temperature\footnote{This is clearly a simplification of open-system dynamics, but it is a useful model and commonly used in quantum thermodynamics \cite{ptherm_aguilar,ptherm_alhambra,ptherm_lostaglio,ptherm_perry,ptherm_quadeer}. Similar dynamics can be recovered from collisional models of thermalisation \cite{ptherm_scarani_collisional}.}. The heat absorbed by the system is given by $Q = \tr[H(\lambda \rho + (1-\lambda)\tau)] - \tr[H\rho]$.}

\vspace{1em}
We call ${\mathcal S}$ the set of all transformations that can be achieved as an arbitrary sequence of unitaries and partial thermalisations, where the Hamiltonian may be changed between one partial thermalisation step and another. The objective, then, is to compose the \textcolor{\myblue}{dual-purpose} qubit operation $\mathcal{M}\in \mathcal{S}$ which maps $\rho_1 \mapsto \eta_1$ and $\rho_2 \mapsto \eta_2$; and assuming this can be done, to optimise its thermodynamic performance.

\section{Feasibility}\label{feasibility}

\begin{figure*}[hbt!]
\includegraphics[height=18em]{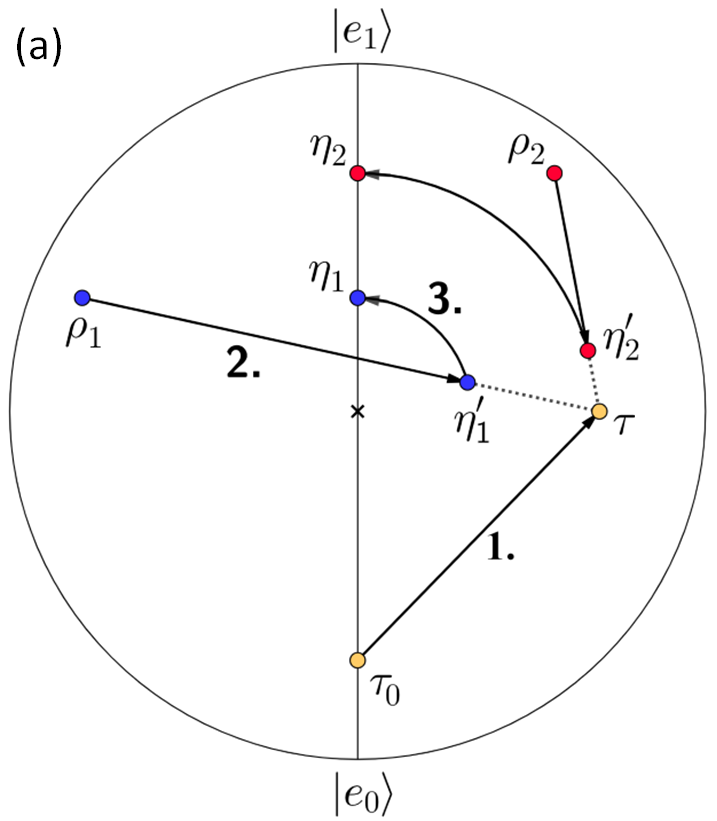}
\quad \quad \quad
\includegraphics[height=18em]{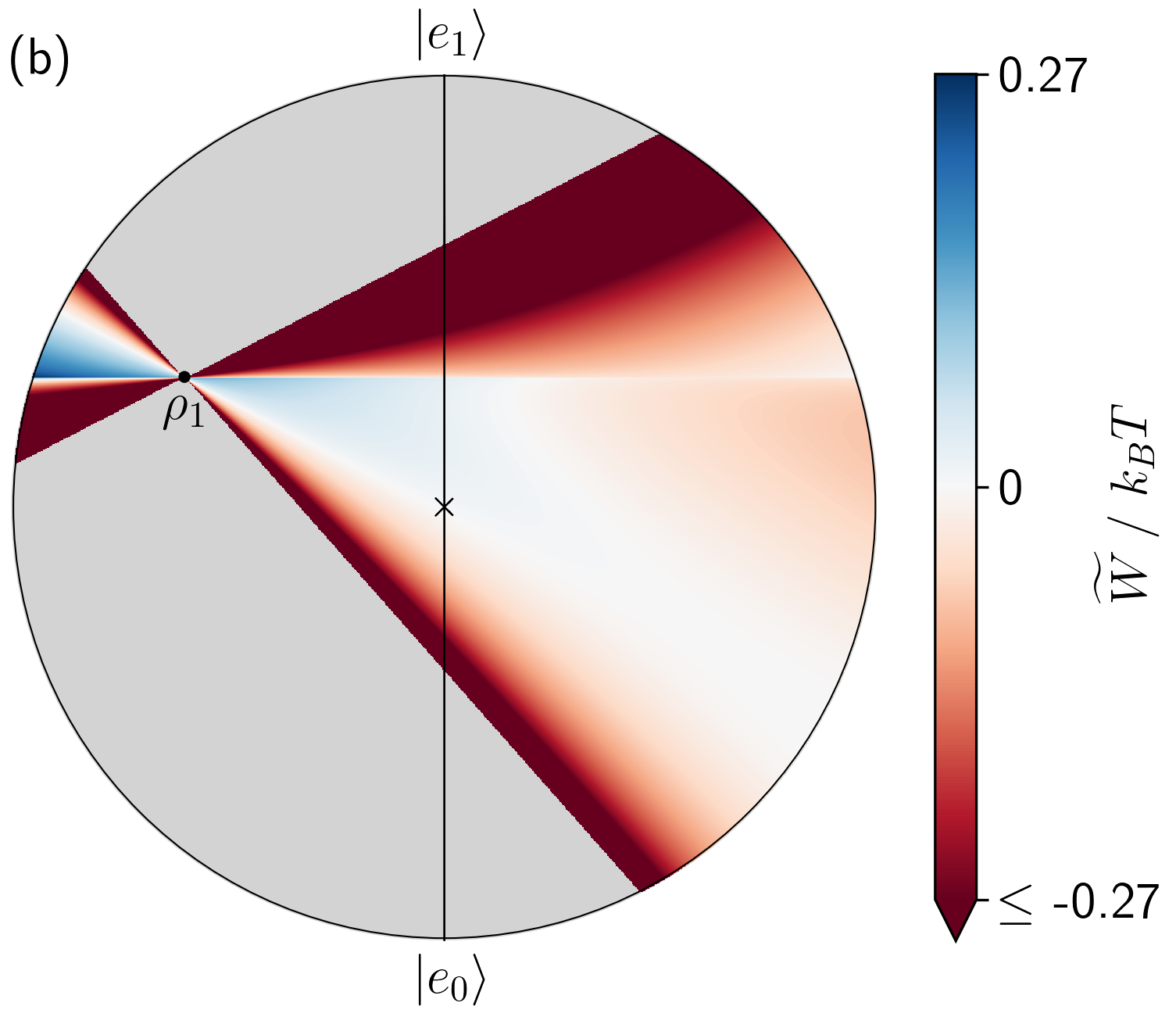}
\centering
\caption{
\textbf{(a)} A work-from-coherence protocol represented in the Bloch sphere. The operation removes coherences in the energy eigenbasis for two possible input states $\rho_1$ and $\rho_2$ while preserving energy level populations, leaving the system in the final state $\eta_1$ or $\eta_2$. \textbf{1.} The qubit Hamiltonian is quickly adjusted, changing the Gibbs state from $\tau_0$ to $\tau$. \textbf{2.} The qubit is allowed to partially thermalise: the initial state - say, $\rho_1$ - is transformed to a mixture with the Gibbs state: $\eta'_1 = \lambda \rho_1 + (1-\lambda) \tau$. \textbf{3.} The Hamiltonian is quickly reset to its original value, and a unitary rotation $U$ is applied, taking the qubit from $\eta'_1$ to the final state $\eta_1$. The values of $\lambda$, $\tau$ and $U$ are uniquely chosen such that the same steps would also transform $\rho_2$ to $\eta_2$. \textbf{(b)} Extractable work plotted as a function of $\rho_2$, in the plane containing $\rho_1, \ket{e_0}$ and $\ket{e_1}$. In the grey region, it is not possible to construct a protocol out of unitaries and partial thermalisations that works for both inputs: in the geometric construction of panel (a), the thermal state $\tau$ would have to lie \textit{outside} the Bloch sphere. This is in contrast with work-from-coherence protocols optimised for a single input: in that case, positive work extraction is possible for any state with coherences \cite{Anders}. $\wtilde$ is obtained by numerically optimising Eq.\eqref{workish} with $p_1{=}p_2{=}\frac{1}{2}$ and $N{=}20$.}
\label{fig:blochspheres}
\end{figure*}

Before turning to thermodynamic considerations, we examine whether it is actually possible to carry out a two-input qubit operation within the framework set out above. As we will find, the answer depends nontrivially on the input and output states. To see this, we first establish that the set $\mathcal{S}$ of possible transformations is identical to the set $\mathcal{S}_1 \subset \mathcal{S}$ of transformations that can be carried out with a \textit{single} partial thermalisation followed by a unitary. Given inputs and outputs for which the transformation is feasible, we are able to explicitly construct a protocol belonging to $\mathcal{S}_1$ (see Fig. \ref{fig:blochspheres}a).

As a starting point, consider a generic qubit map $\mathcal{M}\in\mathcal{S}$ composed as a finite sequence of unitaries $\mathcal{U}$ and partial thermalisations $\mathcal{T}$; something of the form $\mathcal{M} = \mathcal{T}\mathcal{T}\mathcal{U}\mathcal{U}\mathcal{U}\mathcal{T}\mathcal{U}...\mathcal{T}\mathcal{U}$. By combining consecutive unitaries and separating consecutive thermalisations with trivial unitaries $\mathcal{U}=\id$, such an operation can always be rewritten in the form $\mathcal{T}_N\mathcal{U}_N ... \mathcal{T}_2\mathcal{U}_2 \mathcal{T}_1\mathcal{U}_1$. Next we observe that any composite operation of the form $\mathcal{U}^{-1}\mathcal{T}\mathcal{U}$ carries out a mapping identical to a partial thermalisation $\mathcal{T}'$ towards $\tau'=U^\dagger \tau U$. To see this, consider any state $\rho$, where we obtain:
\begin{equation}
\begin{split}
    \mathcal{U}^{-1} \mathcal{T} \mathcal{U} (\rho) &= U^\dagger \big[\lambda [U \rho U^\dagger] +(1-\lambda)\tau\big] U\\
    &= \lambda \rho + (1-\lambda) U^\dagger \tau U\\
    &= \mathcal{T}'(\rho).
\end{split}
\end{equation}
If $\tau$ is the Gibbs state for a Hamiltonian $H$, then $\tau'$ is the Gibbs state for $H'=U^\dagger H U$. In addition, the heat absorbed from the environment is the same for both processes, and equal to the change in internal energy during the partial thermalisation step:
\begin{equation}
\begin{split}
    Q &= \tr[H\,\left(\lambda [U \rho U^\dagger] +(1-\lambda)\tau\right)] - \tr[H [U \rho U^\dagger] ]\\
    &= \tr[H'\,\left(\lambda \rho +(1-\lambda)\tau'\right)] - \tr[H'\rho]\\
    &= Q'.
\end{split}
\end{equation}
For any state, $\mathcal{U}^{-1}\mathcal{T}\mathcal{U}$ and $\mathcal{T}'$ carry out the same transformation at the same energy cost, and so they are for our purposes equivalent. This can be employed to further simplify our ansatz operation. Writing $\mathcal{U}'_i = \mathcal{U}_i ... \mathcal{U}_2 \mathcal{U}_1$, and $\mathcal{T}'_i = {\mathcal{U}'_i}^{-1} \mathcal{T}_i \mathcal{U}'_i$, and using the fact that $\mathcal{U}'_i ({\mathcal{U}'_{i-1}})^{-1} = \mathcal{U}_i$,
\begin{equation}
\begin{split}
    & \mathcal{T}_N\mathcal{U}_N\mathcal{T}_{N-1}\mathcal{U}_{N-1}\,...\,\mathcal{T}_2\mathcal{U}_2 \mathcal{T}_1\mathcal{U}_1\\
    &= \mathcal{U}'_N\left({\mathcal{U}'_N}^{-1} \mathcal{T}_N \mathcal{U}'_N\right)\left({\mathcal{U}'_{N-1}}^{-1} \mathcal{T}_{N-1} \mathcal{U}'_{N-1}\right)...\\
    &\hspace{8em}...\left({\mathcal{U}'_2}^{-1} \mathcal{T}_2 \mathcal{U}'_2\right)\left({\mathcal{U}'_1}^{-1} \mathcal{T}_1 \mathcal{U}'_1\right)\\
    &=\mathcal{U}'_N \,\mathcal{T}'_N\mathcal{T}'_{N-1}\,...\,\mathcal{T}'_2\mathcal{T}'_1.
\end{split}
\end{equation}
So, our original operation $\mathcal{M}$ could be implemented at equal energy cost by a sequence of consecutive partial thermalisations
$\mathcal{T}'_i$ followed by a single unitary $\mathcal{U'}$. While the latter might pose greater practical challenges, the fundamental limitations are unchanged. Hence we can drop the primes and consider protocols of the form $\mathcal{U}\mathcal{T}_N ...\mathcal{T}_2\mathcal{T}_1$ for the rest of the paper, without loss of generality\footnote{Strictly speaking, a quench still occurs between successive partial thermalisations, but its immediate effect on the qubit state is the identity map, which is omitted here for conciseness.}. The next step is to recognise that $\mathcal{T}_N...\mathcal{T}_2\mathcal{T}_1$ can be replaced by a single partial thermalisation. Starting with $N=2$, let $\mathcal{T}_1$ be characterised by $(\lambda_1,\tau_1)$ and $\mathcal{T}_2$ by $(\lambda_2,\tau_2)$. Then for any state $\rho$,
\begin{equation}
    \begin{split}
        \mathcal{T}_2 \mathcal{T}_1 (\rho) &= \lambda_2\left[\lambda_1\rho + (1-\lambda_1)\tau_1\right] + (1-\lambda_2)\tau_2\\
        &= \lambda_2\lambda_1\rho + (1-\lambda_2\lambda_1)\Bigg[\frac{\lambda_2(1-\lambda_1)}{1-\lambda_2\lambda_1}\tau_1 \\
        &\hspace{10em}+ \frac{1-\lambda_2}{1-\lambda_2\lambda_1}\tau_2\Bigg]\\
        &\equiv \lambda^\mathrm{eff}_2 \rho + (1-\lambda^\mathrm{eff}_2)\tau^\mathrm{eff}_2,
    \end{split}
\end{equation}
where we have defined $\lambda^\mathrm{eff}_2 = \lambda_2 \lambda_1$ and $\tau^\mathrm{eff}_2 = \frac{\lambda_2(1-\lambda_1)}{1-\lambda_2\lambda_1}\tau_1 + \frac{1-\lambda_2}{1-\lambda_2\lambda_1}\tau_2$. It is straightforward to check that $\lambda^\mathrm{eff}_2 \in [0,1]$, and that $\tau^\mathrm{eff}_2$ is a convex combination of density operators and is therefore itself a density operator. Hence $\tau^\mathrm{eff}_2$ can be understood to be the thermal state corresponding to a Hamiltonian $-k_\mathrm{B} T (\ln\tau^\mathrm{eff}_2 + \ln Z\, \id)$. So, a single partial thermalisation $\mathcal{T}^\mathrm{eff}_2$ characterised by $(\lambda^\mathrm{eff}_2,\tau^\mathrm{eff}_2)$ carries out the same mapping as $\mathcal{T}_2 \mathcal{T}_1$.

The result extends inductively: if $\mathcal{T}_N...\mathcal{T}_1$ reduces to $\mathcal{T}^\mathrm{eff}_N$, then $\mathcal{T}_{N+1}\mathcal{T}_N...\mathcal{T}_1$ can be written as $\mathcal{T}_{N+1}\mathcal{T}^\mathrm{eff}_N$, which can be reduced to $\mathcal{T}^\mathrm{eff}_{N+1}$, by the above. The mixing parameter and thermal state are given by
\begin{equation}\label{effectives}
    \begin{split}
        \lambda^\mathrm{eff}_N &= \prod_{i=1}^N \lambda_i\\
        \tau^\mathrm{eff}_N &= \frac{\lambda_N(1-\lambda^\mathrm{eff}_{N-1})}{1-\lambda^\mathrm{eff}_N}\tau^\mathrm{eff}_{N-1} + \frac{1-\lambda_N}{1-\lambda^\mathrm{eff}_N}\tau_N
    \end{split}
\end{equation}
where $\lambda^\mathrm{eff}_0{=}\,1$ and $\tau^\mathrm{eff}_0 {=}\, \id$. The upshot of this is that a two-step protocol $\mathcal{U}\mathcal{T}^\mathrm{eff}_N$ carries out the same mapping as the $N{+}1$ step protocol $\mathcal{U}\mathcal{T}_N...\mathcal{T}_1$.

What we have shown is that for every operation $\mathcal{M}\in\mathcal{S}$, we also have $\mathcal{M}\in\mathcal{S}_1$. There is an important caveat: while $\mathcal{U}\mathcal{T}^\mathrm{eff}_N$ executes the same mapping as $\mathcal{U}\mathcal{T}_N...\mathcal{T}_1$, the total heat absorbed during the process will typically differ. As we will see in section \ref{energetics}, it is generally thermodynamically favourable to use $N{\gg}1$ thermalising steps.

We are now ready to decide on the existence of a suitable qubit transformation acting on two generic input states, and give details on how to construct it. Consider an operation $\mathcal{M}\in\mathcal{S}$, with input states $\rho_1,\rho_2$ and outputs $\eta_1,\eta_2$\footnote{We assume a few things about the inputs and outputs. Firstly, that $(\rho_1,\rho_2)\neq(\eta_1,\eta_2)$, so that the operation is not simply the identity. Secondly, that $\eta_1\neq\eta_2$: otherwise, we could simply adjust the Hamiltonian such that the Gibbs state coincides with $\eta_1$, and allow the qubit to completely thermalise. This second assumption also implies that $\rho_1\neq\rho_2$, because CPTP quantum operations are never one-to-many.}. By the above, $\mathcal{M}$ can be written as $\mathcal{U}\mathcal{T}$ where $\mathcal{U}$ is characterised by some  unitary matrix $U$, and $\mathcal{T}$ by a mixing parameter $\lambda\in [0,1]$ and density matrix $\tau$. Then $\lambda,\tau$ and $U$ must satisfy the following:
\begin{equation}\label{simul}
    \begin{cases}
    U [\lambda\rho_1 + (1-\lambda)\tau]U^\dagger &= \eta_1\\
    U [\lambda\rho_2 + (1-\lambda)\tau]U^\dagger &= \eta_2.
    \end{cases}
\end{equation}
Subtracting the second line from the first we obtain
\begin{equation}\label{subtract}
    \lambda U(\rho_1-\rho_2)U^\dagger = \eta_1 - \eta_2,
\end{equation}
and taking the trace norm\footnote{$\norm{\rho}_1 = \tr[\sqrt{\rho\rho^\dagger}]$} of both sides leads to an expression for the mixing parameter:
\begin{equation}\label{lambda}
\lambda= \frac{\norm{\eta_1-\eta_2}_1}{\norm{\rho_1-\rho_2}_1}.
\end{equation}
The requirement that $\lambda\,{\leq}\,1$ necessitates that $\norm{\eta_1-\eta_2}_1 \,{\leq}\, \norm{\rho_2-\rho_1}_1$. Since $\rho_1{-}\rho_2$ is a traceless Hermitian operator on a qubit, it can be diagonalised as $p\ketbra{\psi_+}{\psi_+}\,{-}\,p\ketbra{\psi_-}{\psi_-}$, for some $p\,{>}\,0$, and some orthonormal states $\ket{\psi_+}$ and $\ket{\psi_-}$. Likewise, we can write $\eta_1{-}\eta_2 = q\ketbra{\phi_+}{\phi_+}-q\ketbra{\phi_-}{\phi_-}$: note that $\norm{\rho_1\,{-}\,\rho_2}_1 = 2p$ and $\norm{\eta_1\,{-}\,\eta_2}_1 = 2q$. Substituting these expressions into Eq. \eqref{subtract} leads to a formula for $U$:
\begin{equation}\label{U}
U = \ketbra{\phi_+}{\psi_+}+\ketbra{\phi_-}{\psi_-}.
\end{equation}
Now that $\lambda$ and $U$ have been determined in terms of $\rho_1,\rho_2,\eta_1$ and $\eta_2$, a rearrangement of the first line\footnote{Doing the same for the second line (or any affine combination of the two lines) would give an equivalent expression.} of Eq. \eqref{simul} yields an expression for $\tau$,
\begin{equation}\label{tau}
    \tau = \frac{1}{1-\lambda}\left[U^\dagger \eta_1 U - \lambda \rho_1\right].
\end{equation}

Equation \eqref{tau} by itself guarantees that $\tau$ is a Hermitian operator with unit trace. However, we also require $\tau\,{\geq}\,0$ in order that $\tau$ is a well-defined density matrix.

So, given an operation specified by inputs $\rho_1,\rho_2$ and outputs $\eta_1,\eta_2$, it is straightforward to determine whether the operation belongs in $\mathcal{S}_1$: we can evaluate $\lambda,U$ and $\tau$ using equations (\ref{lambda} - \ref{tau}), and check that $\lambda\,{\leq}\,1$ and $ \tau\,{\geq}\,0$. If both these conditions hold, then the above construction provides an explicit protocol $\mathcal{U}\mathcal{T}$ to carry out the operation (see Fig. \ref{fig:blochspheres}a). Moreover, the operation can also be implemented using an extended protocol $\mathcal{U}\mathcal{T}_N...\mathcal{T}_1 \in \mathcal{S}_N$ for which $\lambda^\mathrm{eff}_N$ and $\tau^\mathrm{eff}_N$ as given in \eqref{effectives} are equal to $\lambda$ and $\tau$ (\ref{lambda},\ref{tau}). Conversely, if $\lambda\,{>}\,1$ or if $\tau$ has a negative eigenvalue, then the \textcolor{\myblue}{dual-purpose} operation cannot be composed as a sequence of unitaries and partial thermalisations, no matter how many steps are used\footnote{Not all CPTP channels can be composed in this way: the set of feasible operations can be expanded through the use of an ancilla system. To operate in a cycle, the ancilla's state must be reset, erasing thermodynamically relevant system-ancilla correlations \cite{maroney,delrio}. For the simplicity of analysis, we here consider only those qubit operations which can be implemented without the use of ancillas.}.

\section{Work Extraction}\label{energetics}

The decomposition of a qubit operation into unitaries and partial thermalisations is not unique - and the various ways it can be done generally involve differing net transfers of work and heat.
This leads to the question of optimal implementation: for a given set of inputs and outputs, how can the operation be carried out so as to yield as much work as possible (or expend as little as possible)? And how does the optimum compare against the fundamental bound set by the second law?

We approach this problem by first deriving an expression for the average work extraction for a protocol involving $N$ partial thermalisation steps. This is employed in section \ref{wfcsection} to numerically solve for optimum \textit{work from coherence} protocols for two inputs, allowing for comparison against the reversible single-input protocol in \cite{Anders}. In section \ref{boundsection} we derive a general upper bound on work extraction for two-input operations, opening up an interpretation of why thermodynamic reversibility cannot be achieved.

We choose as a figure of merit the average work yield, $\wk = p_1 W_1 + p_2 W_2$, where $p_1$ is the probability that the input is $\rho_1$, and $W_1$ is the work extracted in transforming $\rho_1$ to $\eta_1$  (and likewise for the second input). Since the overall change in the qubit's expected energy is path-independent, then by the first law ($\Delta \overline{U} = \Q - \wk$), maximising work extraction amounts to the same thing as maximising the heat drawn from the environment.

First, let us evaluate the heat for a single partial thermalisation, with mixing parameter $\lambda$ and thermal state $\tau$, acting on a state $\rho$. The heat absorbed is equal to the resulting difference in the qubit's expected energy:
\begin{equation}
\begin{split}
    Q &= \tr[H(\lambda\rho + (1-\lambda)\tau)] - \tr[H\rho]\\
    &= -k_\mathrm{B} T (1-\lambda) \tr[(\tau-\rho)\ln\tau].
    \end{split}
\end{equation}
In the second line, we used that $H = -k_\mathrm{B} T [\ln(\tau) + \ln Z\,\id]$. Now, consider a protocol $\mathcal{U}\mathcal{T}_N ... \mathcal{T}_1$:
The average heat drawn from the environment is given by a sum over the $N$ partial thermalisation steps, averaged over the input states (the unitary step involves no heat transfer). Letting $\rho_n^i = \mathcal{T}_{i-1} ... \mathcal{T}_1 (\rho_n)$ denote the transformed state of $\rho_n$ prior to the $i^\mathrm{th}$ partial thermalisation, for $n=1,2$, we have:
\begin{equation}\label{finalheat}
    \begin{split}
        \Q &= -k_\mathrm{B} T \sum_{n=1}^2 p_n \sum_{i=1}^N (1{-}\lambda_i) \tr[\left(\tau_i-\rho_n^i\right)\ln\tau_i]\\
        &= -k_\mathrm{B} T \sum_{i=1}^N (1{-}\lambda_i) \tr[\left(\tau_i-\overline{\rho}_i\right)\ln\tau_i]\\
        &= k_\mathrm{B} T \sum_{i=1}^N (1{-}\lambda_i) \left[S(\tau_i) - S(\overline{\rho}_i) - S(\overline{\rho}_i||\tau_i)\right].
    \end{split}
\end{equation}
where $\overline{\rho}_i \,\,{=}\,\, \mathcal{T}_{i-1} ... \mathcal{T}_1 (\overline{\rho}) \,\,{\equiv}\,\, p_1 \rho_1^i \,\,{+}\,\, p_2 \rho_2^i$ is the average state before the $i^\mathrm{th}$ partial thermalisation, and $S(\rho||\sigma) = \tr[\rho\ln\rho - \rho\ln\sigma]$ is the quantum relative entropy.
Invoking the first law, the work yield is given by subtracting the overall change of expected energy from \eqref{finalheat}, to give:
\begin{equation}\label{workish}
\begin{split}
    \wk = k_\mathrm{B} T &\sum_{i=1}^N (1{-}\lambda_i) \left[S(\tau_i) - S(\overline{\rho}_i) - S(\overline{\rho}_i||\tau_i)\right]\\
        -&\tr[H_0 (\overline{\eta}-\overline{\rho})].
\end{split}
\end{equation}
The above is straightforward to evaluate given the parameters $\lambda_i,\tau_i$. However, we are ultimately interested in the maximum possible work extraction, $\wtilde$, over all suitable choices of protocol:
\begin{equation}\label{tilde}
    \wtilde = \lim_{N\to\infty} \max_{\{\lambda_i\},\{\tau_i\}} \wk,
\end{equation}
where the maximisation\footnote{The optimum work extraction must increase monotonically with the number of partial thermalisation steps $N$: for any $N$-step protocol we can construct an $N{+}1$ step protocol for the same operation with identical work yield by decomposing one of its partial thermalisation steps $\mathcal{T}_i = (\lambda_i,\tau_i)$ into two consecutive steps $\mathcal{T}^{(1)}_i\,{=}\,\mathcal{T}^{(2)}_i\,{=}\,(\sqrt{\lambda_i},\tau_i)$.} is subject to the constraints that $\lambda^\mathrm{eff}_N \,{=}\, \lambda$ and $\tau^\mathrm{eff}_N \,{=}\, \tau$ as given in equations (\ref{effectives},\ref{lambda},\ref{tau}), ensuring that the protocol maps $\rho_1\mapsto\eta_1$ and $\rho_2\mapsto\eta_2$. Note that the expression for $\wk$ \eqref{workish} only depends on the trajectory of the average state $\overline{\rho}$: the individual inputs and outputs $\rho_1,\rho_2,\eta_1$ and $\eta_2$ enter Eq. \eqref{tilde} implicitly through the optimisation constraints. 

Recall that some operations are unfeasible because any suitable protocol would involve thermalisation towards a state $\tau_i$ with a negative eigenvalue. For operations at the \textit{boundary} of feasibility, $\tau_i$ must approach a pure state, with a zero eigenvalue. But in this limit, the relative entropy term $S(\overline{\rho}_i||\tau_i)$ appearing in \eqref{workish} - and consequently the work cost of the operation - diverge towards $+\infty$.

While \eqref{tilde} does not admit a simple closed form, it does provide a starting point for numerical investigations, and will later provide the basis for an analytic upper bound on work extraction, \eqref{finalbound}. We effectively already have a lower bound on $\wtilde$ stemming from the protocol derived at the end of section \ref{feasibility}: the expressions for $\lambda$ and $\tau$ (\ref{lambda},\ref{tau}) can be directly substituted into \eqref{workish}.

\begin{figure*}[hbt!]
    \includegraphics[height=16em]{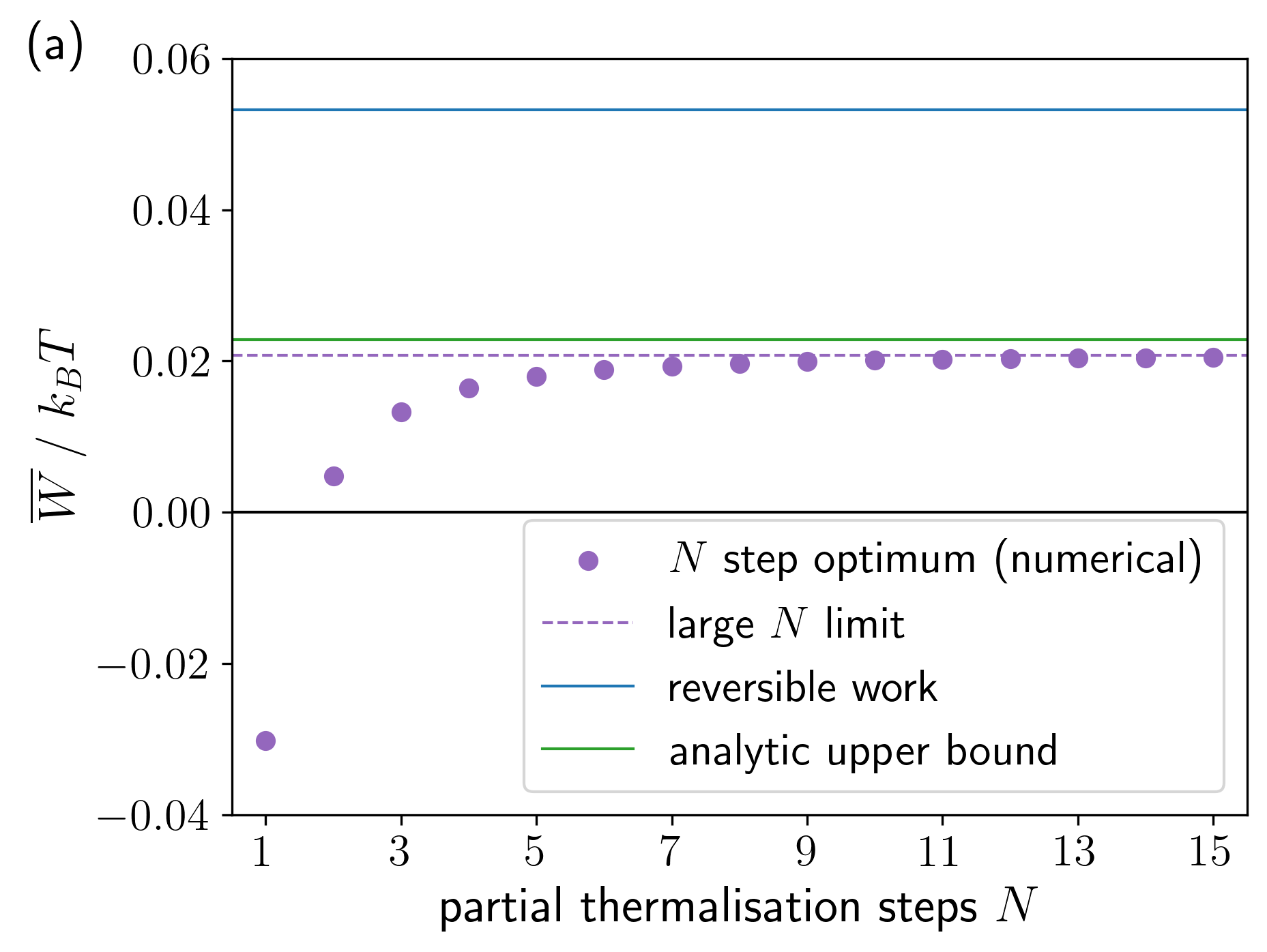}
    \quad
    \includegraphics[height=16em]{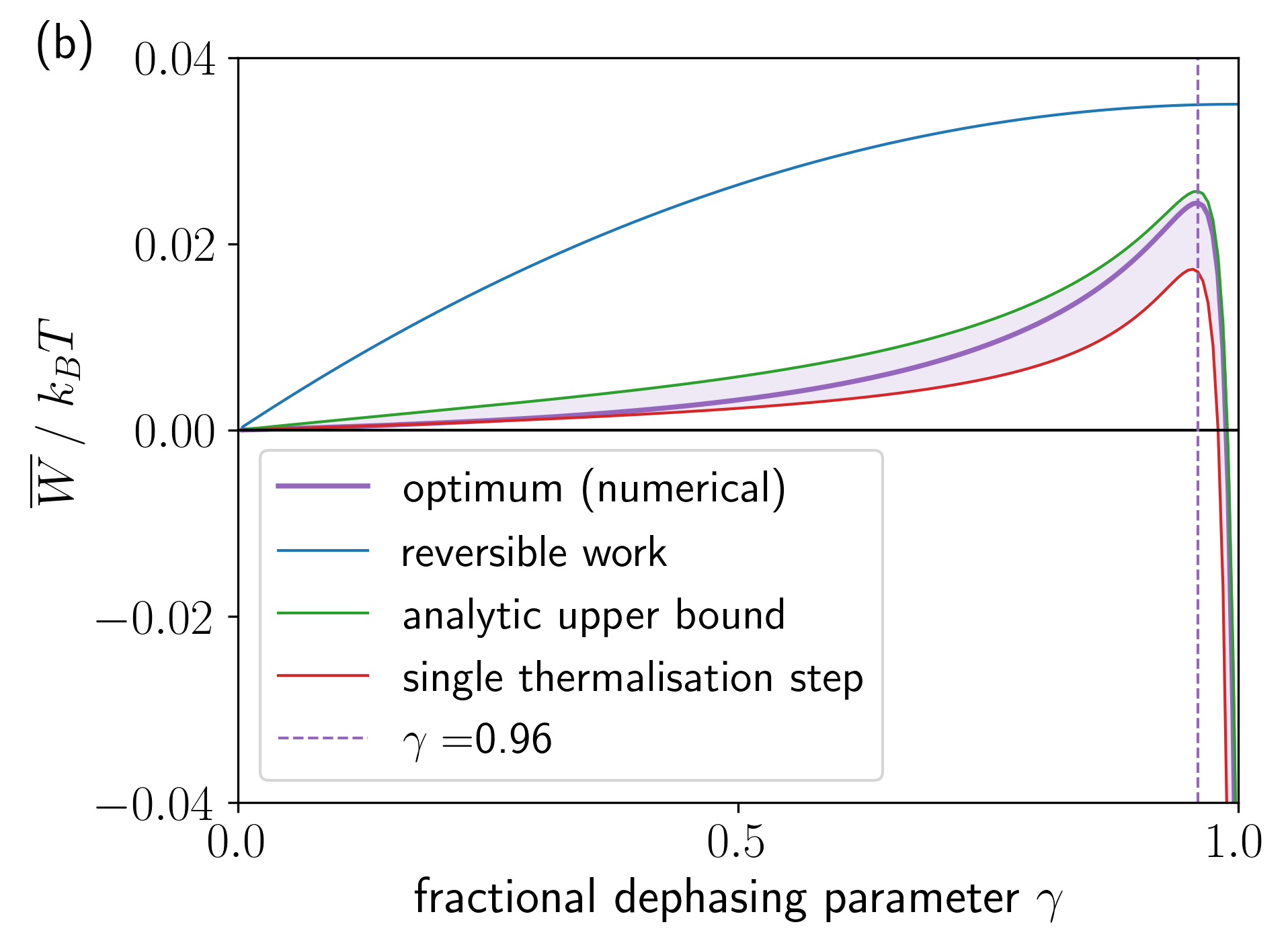}
    \caption{
    \textbf{(a)} Coherence work extraction for protocols involving increasing numbers of partial thermalisation steps $N$, obtained via numerical optimisation (purple scatter). Thermodynamic performance rapidly improves with $N$, converging towards an upper limit (dotted line) which is still significantly lower than the work extraction for a thermodynamically reversible operation, given by the free energy reduction $k_\mathrm{B} T [S(\overline{\eta})-S(\overline{\rho})]$ (blue line). The improved analytic bound, Eq. \eqref{finalbound} (green line) provides a closer estimate of the \textcolor{\myblue}{dual-purpose} work extraction, accounting for the constraint of preserving distinguishability between the two input states.
    The scatter was obtained by numerically maximising Eq. \eqref{workish} over $\tau_1,...,\tau_N$, for input states with Bloch vectors $\rho_1 = (0.735,0.273,-0.286)$ and $\rho_2 = (-0.496,-0.470,-0.294)$, and outputs $\eta_1 = (0,0,-0.286)$ and $\eta_2 = (0,0,-0.294)$, with $p_1 {=} p_2 {=} \frac{1}{2}$. The initial Hamiltonian is taken as $H_0 = E_0 \sigma_z$. \textbf{(b)} Work extraction for an operation which \textit{partially} removes coherences in the energy eigenbasis (purple curve). Off-diagonal elements $\bra{e_m}\rho\ket{e_n}$ are scaled by a factor $1{-}\,\gamma$, such that $\gamma=0$ leaves the state unchanged and $\gamma=1$ corresponds to complete dephasing. Surprisingly, for this pair of input states, maximum work extraction is \textit{not} achieved for complete dephasing, where the reduction of free energy (blue curve) is greatest: in fact the maximum is at $\gamma\approx 0.96$ (dotted line). The analytic bound \eqref{finalbound} (green curve) better captures the dependence of work extraction on $\gamma$. On the other hand, the single step work extraction (red curve) provides a lower bound on the optimum - and since $\lambda$ \eqref{lambda} and $\tau$ \eqref{tau} are uniquely determined in this case, we can constrain $\wtilde$ to the shaded window in the figure without performing any numerical optimisation.
    For this plot, the input Bloch vectors were $\rho_1 = (0.249, 0.183, 0.494)$ and $\rho_2 = (-0.044, -0.640,  0.508)$, and the outputs $\eta_1 = (0.249(1{-}\,\gamma), 0.183(1{-}\,\gamma), 0.494)$ and $\eta_2 = (-0.044(1{-}\,\gamma), -0.640(1{-}\,\gamma),  0.508)$, with $p_1 {=} p_2 {=} \frac{1}{2}$. The optimum two-input work output was computed by maximising \eqref{workish} using $N{=}20$ steps: as can be seen in panel (a), this is sufficient for the work extraction to converge with reasonable precision. For both panels, the input states were generated randomly and post-selected to illustrate the sign change for work extraction with $2+$ thermalisation steps in panel (a); and to show the thermodynamic advantage of incomplete dephasing in panel (b).
    }
    \label{workplots}
\end{figure*}
 
\subsection{Work from coherence}\label{wfcsection}

In the previous subsection we derived a formula for the average work yield of a two-input qubit operation \eqref{workish}. In order to better understand its properties, let us apply it to a specific operation: work extraction from coherence.

By this we mean a process of work extraction which leaves unchanged the state's diagonal matrix elements $\bra{e_n}\rho\ket{e_n}$ in the eigenbasis of the initial Hamiltonian $H_0$, and which involves a single heat bath at fixed temperature $T$ and a cyclic variation of the Hamiltonian, returning it to $H_0$ at the end of the process \cite{Anders,Smith,Mohammady}. In this context, coherences - that is, off-diagonal elements $\bra{e_m}\rho\ket{e_n}$ - represent a thermodynamic resource. A state $\rho$ with nonzero coherences has lower von Neumann entropy than the corresponding \textit{decohered} state $\eta = \mathrm{diag}(\rho) := \ketbra{e_0}{e_0}\rho\ketbra{e_0}{e_0} + \ketbra{e_1}{e_1}\rho\ketbra{e_1}{e_1}$. Since the expected energy is unchanged, i.e. $\tr[H_0 ({\eta}-{\rho})]\,{=}\,0$, transforming $\rho$ to $\eta$ results in a reduction in free energy $\Delta F =  k_\mathrm{B} T [S(\eta)-S(\rho)]\leq 0$. Consequently, if carried out reversibly, the operation can extract positive work, such as in the protocol developed in Ref. \cite{Anders} for a single input state.

Still, the question remains how closely a \textit{\textcolor{\myblue}{dual-purpose}} work from coherence protocol can approach the reversible work yield. We consider an operation with two inputs $\rho_{1,2}$, and completely decohered outputs\footnote{By linearity, $\overline{\eta} =  \mathrm{diag}(\overline{\rho})$.}:
\begin{equation}\label{dephasing}
\eta_{1,2} = \mathrm{diag}(\rho_{1,2}).
\end{equation}
 Fig. (\ref{fig:blochspheres}a) represents a protocol which carries out this mapping using a single partial thermalisation step. If the process could be carried out in a thermodynamically reversible way, it would extract work equal to the free energy reduction $-\Delta \overline{F} = k_\mathrm{B} T [S(\overline{\eta})-S(\overline{\rho})]\geq 0$, on average over the inputs.

On the other hand, the actual maximum work extraction can be estimated numerically using Eq. \eqref{workish}. Fixing the states $\rho_1,\rho_2$ and the number of partial thermalisation steps $N$, the work output $\wk$ can be maximised as a function of the thermal states $\tau_1,...,\tau_{N-1}$ and mixing parameters $\lambda_1,...,\lambda_{N-1}$\footnote{For the final partial thermalisation, $\lambda_N$ and $\tau_N$ are fixed by the constraint that $\lambda^\mathrm{eff}_N = \lambda$ and $\tau^\mathrm{eff}_N = \tau$.}. If $N$ is chosen sufficiently large, the resulting value of $\wk$ will be a close approximation to the overall optimum $\wtilde$. Fig. (\ref{workplots}a) shows convergent behaviour of $\wk$ with increasing $N$. Typically, $20$ thermalisation steps are sufficient for the work yield to converge within a $1\%$ margin of its limiting value. As an initial guess for optimisation, we took that for all $i$, $\tau_i = \tau$ as given in Eq.\eqref{tau}. It was assumed\footnote{While this is a nontrivial assumption for finite $N$, it should make no difference in the limit as $N\to\infty$. Any partial thermalisation step $(\lambda,\tau)$ can be approximated by $n$ consecutive steps $\left(\left(\frac{\norm{\eta_1-\eta_2}_1}{\norm{\rho_1-\rho_2}_1}\right)^{\frac{1}{N}},\tau\right)$, where $n$ is chosen such that $\left(\frac{\norm{\eta_1-\eta_2}_1}{\norm{\rho_1-\rho_2}_1}\right)^{\frac{n}{N}}\approx \lambda$.} throughout that $\lambda_i = \left(\frac{\norm{\eta_1-\eta_2}_1}{\norm{\rho_1-\rho_2}_1}\right)^{\frac{1}{N}}$.

The resulting numerical approximation to $\wtilde$ is conspicuously lower than the change in free energy. In fact, for many pairs of input states, work extraction is not possible at all. Uniform sampling of the Bloch sphere reveals that the operation \eqref{dephasing} is only \textit{feasible} for $62\%$ of pairs of input states: the remainder violate the conditions established at the end of section \ref{feasibility}, and there is no way to carry out the intended mapping for both inputs using unitaries and partial thermalisations. Positive work extraction is only possible for $10\%$ of the feasible input pairs with $p_1 = p_2 = \tfrac{1}{2}$. The heatmap in Fig. (\ref{fig:blochspheres}b) gives an indication of the relative scale of these regimes: unfeasible pairs are marked in grey, and ones where work can be extracted are marked in blue. For the majority of states, it is possible to remove coherences while preserving the energy distribution, but it costs work to do so - these pairs lie in the red regions.

This is a stark departure from single input work-from-coherence, which can be performed reversibly for any state, and can \emph{always} be used to extract work. Evidently, the compromises necessary for a \textcolor{\myblue}{dual-purpose} operation impose a severe energetic penalty. In section \ref{boundsection} we will investigate the thermodynamic explanation for this penalty, and derive an improved bound on work extraction, Eq. \eqref{finalbound} (plotted as the green lines in Fig. \ref{workplots}).

In the meantime there is an interesting caveat to explore. Up to this point we have considered an operation which completely removes energy-basis coherences. However, in a broader sense, work from coherence can include any operation which leaves the energy level populations unchanged - that is, one for which $\mathrm{diag}(\eta_{1,2})$ = $\mathrm{diag}(\rho_{1,2})$. This could mean simply scaling the coherences by some factor $1-\gamma\in[0,1]$, such that $\gamma=0$ leaves the state unchanged and $\gamma=1$ corresponds to complete removal of coherences:
\begin{equation}
    \eta_{1,2} = (1-\gamma) \rho_{1,2} + \gamma\,\mathrm{diag}(\rho_{1,2}).
\end{equation}
The work yield for an operation of this type is plotted as a function of $\gamma$ in Fig. (\ref{workplots}b). Surprisingly, maximum work extraction is \textit{not} necessarily achieved by completely removing coherences -- even though that would result in a greater reduction in free energy. For some pairs of inputs, such as those in Fig. (\ref{workplots}b), a partial decoherence operation may extract positive work where the full removal of coherences would cost work to perform, or even where full removal of coherences is not feasible.

\subsection{Bound on extractable work}\label{boundsection}

For operations with a single input and output state, thermodynamic reversibility can be recovered in the limit of large $N$: crucially, the system's state must remain infinitesimally close to thermal equilibrium whenever it is in contact with the environment \cite{Anders_Giovannetti}. This is the familiar quasistatic limit. Why can't we do the same for a multipurpose operation by ensuring that the average state undergoes a quasistatic evolution?

Consider the term $\sum_{i=1}^N (1-\lambda_i)[S(\tau_i)-S(\overline{\rho}_i)]$ appearing in the equation for $\wk$ \eqref{workish}.
Due to the concavity of the von Neumann entropy, for all $i$,
\begin{equation}
\begin{split}
    (1{-}\lambda_i)[S(\tau_i)- S(\overline{\rho}_i)] &= [\lambda_i S(\overline{\rho}_i) + (1{-}\lambda_i)S(\tau_i)]\\
    & \hspace{8em}- S(\overline{\rho_i})\\
    &\leq S(\lambda_i\overline{\rho}_i + (1{-}\lambda_i)\tau_i)\\
    & \hspace{8em}- S(\overline{\rho_i})\\
    &= S(\overline{\rho}_{i+1}) - S(\overline{\rho}_i).
\end{split}
\end{equation}
So, the sum over $i$ is bounded by the following:
\begin{equation}
\begin{split}
    \sum_{i=1}^N (1-\lambda_i)\left[S(\tau_i)-S(\overline{\rho}_i)\right] &\leq \sum_{i=1}^N \left[S(\overline{\rho}_{i+1}) - S(\overline{\rho}_i)\right]\\
    &= S(\overline{\eta}) - S(\overline{\rho}),
\end{split}
\end{equation}
where we have used that the average state prior to the first partial thermalisation step is equal to the average input state, $\overline{\rho}_1 = \overline{\rho}$; and that following the final partial thermalisation the state is related by a unitary to the average output, $\overline{\rho}_{N+1} = U^\dagger\overline{\eta} U$, where $U$ is given in Eq. \eqref{U}.
So, the overall work extraction is bounded by
\begin{equation}\label{lag}
\begin{split}
    \wtilde &\leq k_\mathrm{B} T [S(\overline{\eta}) - S(\overline{\rho})]-\tr[H_0(\overline{\eta}-\overline{\rho})]\\
    &\hspace{2em} - k_\mathrm{B} T \sum_{i=1}^N (1-\lambda_i) S(\overline{\rho}_i||\tau_i).
\end{split}
\end{equation}

\begin{figure}[t!]
    \centering
    \includegraphics[width=0.7\columnwidth]{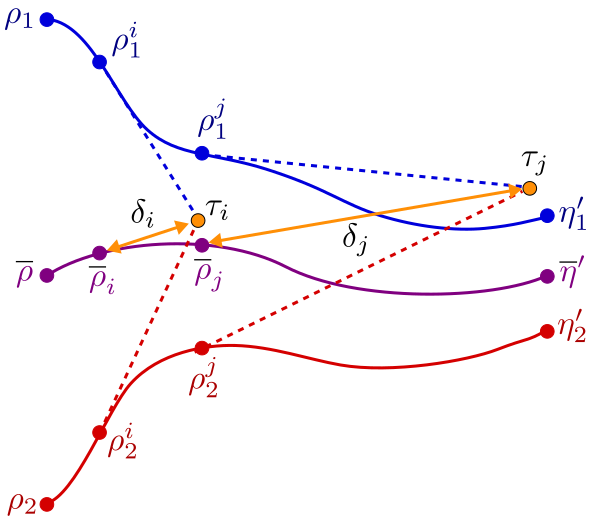}
    \caption{State trajectories during a sequence of many partial thermalisation steps, approximating a smooth curve. In the $i^\mathrm{th}$ step, the qubit state ($\rho^i_1$ or $\rho^i_2$) evolves a small fraction of the distance towards the thermal state $\tau_i$, which varies throughout the protocol: the tangents to the states' trajectories point towards $\tau_i$. The lag $\delta_i = \norm{\overline{\rho}_i-\tau_i}$ of the average state $\overline{\rho}_i$ behind the thermal state $\tau_i$ leads to irreversible entropy production: the process can only approach thermodynamic reversibility if $\delta_i$ can be made arbitrarily small for \emph{every} partial thermalisation step \eqref{lag}. However, if $\delta_i$ is made too small, the distance between the two input states $\norm{\rho^i_1-\rho^i_2}$ will decay much faster than the remaining distance $\norm{\overline{\rho}_i - \overline{\eta}'}$ between the present average state $\overline{\rho}_i$ and the target average state $\overline{\eta}'$ at the end of the process. If $\delta_i$ is too small during one thermalising step, it must be compensated for with a larger $\delta_j$ in another step, in order to reach the target output states while preserving the distinction between $\rho_1$ and $\rho_2$. The average lag $\delta$ throughout the process is lower-bounded, meaning that entropy production and the resulting irreversible dissipation of free energy cannot be eliminated \eqref{finalbound}.}
    \label{sketch}
\end{figure}

The first two terms in \eqref{lag} represent the free energy reduction $-\Delta \overline{F}.$
The final term is a contribution to irreversible entropy production (wasted free energy) $k_\mathrm{B} T\, \Sigma = -\Delta \overline{F} - \widetilde{W}$, resulting from \textit{lag} of the average state behind the thermal state during partial thermalisations. Similar lag terms appear in other works but are usually associated with sub-optimal protocols, for example due to inaccurate knowledge of the initial state, limited control, or finite speed precluding ideal quasistatic processes \cite{Mohammady,Kawai,Campisi,Kolchinsky}. Here we will show that the lag is necessary to preserve dependence on the initial state (see Fig. \ref{sketch}), and the resulting energy penalty is unavoidable for \textcolor{\myblue}{dual-purpose} operations. To do so we employ Pinsker's inequality $S(\rho||\sigma)\geq \frac{1}{2}{\norm{\sigma-\rho}_1}^2$ \cite{ohya} to relate the quantum relative entropy to the trace distance, and thereby bound the entropy production in \eqref{lag} by:
\begin{equation}\label{pinsker}
\begin{split}
    \Sigma &\geq \frac{1}{2} \sum_{i=1}^N (1-\lambda_i) {\norm{\tau_i-\overline{\rho}_i}_1}^2\\
    &= \frac{1}{2} \sum_{i=1}^N \frac{1}{1-\lambda_i} {\norm{\overline{\rho}_{i+1}-\overline{\rho}_i}_1}^2,
\end{split}
\end{equation}
where the second line was obtained by substituting $\overline{\rho}_{i+1} = \lambda_i \overline{\rho}_i + (1-\lambda_i)\tau_i$. At a glance it might appear that the above can be made arbitrarily small. For the average step, $\norm{\overline{\rho}_{i+1}-\overline{\rho}_i}_1$  need not exceed $\frac{1}{N}\norm{U^\dagger\overline{\eta}U - \overline{\rho}}_1$: so the sum of squares in \eqref{pinsker} might vanish in the limit of large $N$, provided that $\frac{1}{1-\lambda_i}$ remains bounded.
However, each partial thermalisation step reduces the distance $\norm{\rho^i_1 - \rho^i_2}$ between the input states by a factor $\lambda_i$ - so the geometric mean of $\lambda_i$ must equal $\left(\frac{\norm{\eta_1-\eta_2}_1}{\norm{\rho_1-\rho_2}_1}\right)^\frac{1}{N}$ (see Eq. \ref{lambda}). As a result, $\frac{1}{1-\lambda_i}$ does indeed diverge for large $N$. Put differently, while $\norm{\overline{\rho}_{i+1}-\overline{\rho}_i}_1$ can vanish for all steps, the distance between the average state and the thermal state $\norm{\tau_i-\overline{\rho}_i}_1 = \frac{1}{1-\lambda_i} \norm{\overline{\rho}_{i+1}-\overline{\rho}_i}_1$ cannot - and this is ultimately the thermodynamically relevant one. Figure \eqref{sketch} illustrates the dilemma of minimising irreversible dissipation while preserving distinguishability between the individual output states.

Applying standard geometric inequalities to Eq. \eqref{pinsker} (appendix \ref{appx}), the extractable work \eqref{lag} can be upper bounded in terms of the input and output states:
\begin{equation}\label{finalbound}
 \widetilde{W} \leq -\Delta \overline{F} - \frac{k_\mathrm{B} T}{2} \frac{{\norm{\,U^\dagger\overline{\eta}\, U\, -\overline{\rho}\,}_1}^2}{\ln(\frac{\norm{\rho_1-\rho_2}_1}{\norm{\eta_1-\eta_2}_1})},
\end{equation}
where the unitary $U$ is given by equation \eqref{U}. The final term, representing unavoidable dissipation, fits well with the intuitive picture presented in Fig. \eqref{sketch}. Looking at the numerator, the dissipation is greater for operations which require the average state $\overline{\rho}$ to be moved a greater distance through the Hilbert space to $U^\dagger \overline{\eta} U$ during the thermalising portion of the protocol. From the denominator, we see that it is costly to preserve a relatively large trace distance between the individual outputs $\eta_1$ and $\eta_2$.

On the other hand, an operation can be made thermodynamically reversible only if the dissipation term \eqref{finalbound} vanishes. Setting the numerator to zero implies that the average output is related to the average input by a unitary\footnote{Moreover, it must be the \textit{same} unitary $U$ \eqref{U} that maps the eigenbasis of $\rho_1-\rho_2$ to that of $\eta_1-\eta_2$. However, doesn't imply that the overall operation is purely unitary, in that $\eta_{1,2} = U \rho_{1,2} U^\dagger$. For example, the operation could remain thermodynamically reversible if the qubit underwent partial thermalisation towards a thermal state $\tau$ identical to the average state $\overline{\rho}$}. Looking at the denominator, dissipation asymptotically vanishes as $\norm{\eta_1-\eta_2}_1\to 0$, with reversibility recovered in the limiting case where both inputs are mapped to the same output. Most prominently, Landauer erasure falls into this latter category, resetting the system from any initial state to a standard $\ketbra{0}{0}$ state. The exceptional cases where the wasted free energy may vanish are in line with those identified by Bedingham and Maroney \cite{maroney}. However, we are able to go further by specifying a procedure to carry out the operation, as well as methods to optimise for maximum work extraction.

The irreversible work cost implied by \eqref{finalbound} does \emph{not} disappear as the probabilities $p_1$ or $p_2$ approach zero. In fact we see a discontinuous jump in $\widetilde{W}$ between the single-input operation where input state $\rho_1$ occurs with certainty, and the \textcolor{\myblue}{dual-purpose} operation where $\rho_1$ occurs with probability $1-\varepsilon$ and $\rho_2$ with probability $\varepsilon$.

While \eqref{finalbound} is not a tight bound, it gives a significantly closer estimate to the optimum work extraction $\wtilde$ than the free energy reduction alone. This is evident in both panels of Fig. \eqref{workplots}, where the improved bound is plotted alongside the reversible work $-\Delta \overline{F}$, as well as numerically optimised protocols.

In practice, the need to avoid complete thermalisation means that thermal contact with the environment must be limited in duration. In a simple model, we might treat the qubit as approaching the Gibbs state at an exponential rate with a fixed timescale $t_\mathrm{th}$, such that for each partial thermalisation step, the mixing parameter $\lambda_i$ can be expressed in terms of the duration of the step $t_i$, as $\lambda_i = \exp(-\frac{t_i}{t_\mathrm{th}})$.

In that case, the total duration $t_\mathrm{tot} = \sum_{i=1}^N t_i$ of thermal contact is determined by the reduction in trace distance between the inputs:
\begin{equation}\label{time}
\begin{split}
    \frac{t_\mathrm{tot}}{t_\mathrm{th}} &= -\ln \prod_{i=1}^N \lambda_i
    = \ln(\frac{\norm{\rho_1 - \rho_2}_1}{\norm{\eta_1 - \eta_2}_1}).
\end{split}
\end{equation}
This in turn means that the irreversible dissipation of free energy is inversely-proportional to $t_\mathrm{tot}$. From \eqref{finalbound}, we have:
\begin{equation}
    -\Delta\overline{F}-\widetilde{W} \geq \frac{k_\mathrm{B} T}{2} \frac{t_\mathrm{th}}{t_\mathrm{tot}}{\norm{U^\dagger \overline{\eta} U - \overline{\rho}}_1}^2.
\end{equation}
Dissipation could vanish in the limit of large $t_\mathrm{tot}$, but the trace distance between the outputs $\norm{\eta_1 - \eta_2}_1$ would vanish too, as can be seen from eq.\eqref{time}. This means that operations with more than one output \textit{cannot} be carried out arbitrarily slowly, even though it would be thermodynamically favourable to do so.

\section{Conclusion and outlook}

In the context of information processing, it is necessary to consider physical processes which produce different final states depending on how the system is initialised. We have examined the thermodynamics of a simple case: qubit transformations with two possible input states. For what might appear a straightforward extension beyond single-input processes, the additional constraints have drastic consequences.

In section \ref{feasibility} we characterised the class of two-input qubit operations which can be implemented through a combination of unitary and partial thermalisation steps, and determined a three step procedure for carrying them out.
In section \ref{energetics}, we found that even for thermodynamically optimal protocols, a considerable fraction of the free energy change must be irreversibly dissipated.
This is attributed to the requirement that the system is significantly out of equilibrium during thermal contact with the environment. By quantifying the resultant entropy production, we derived an improved upper bound on the extractable work (Eq. \ref{finalbound}), and related it to an effective minimum speed limit.

Naturally, there arises the question of extending the present quantitative results to \textcolor{\myblue}{general $n$-input multipurpose} operations. We expect this extension to be nontrivial, not least because for any three states $\rho_1,\rho_2,\rho_3$ of a $d$-dimensional system, the generalised angle $\frac{\tr[(\rho_1-\rho_2)(\rho_1-\rho_3)]}{\tr[\rho_1-\rho_2]\tr[\rho_1-\rho_3]}$ is preserved under both unitaries and partial thermalisations. If this angle were to differ between a triplet of input states and the corresponding outputs, the transformation would be impossible without expanding the set of primitive operations. For $3{+}$ dimensional systems, this could be approached by incorporating \textit{partial level thermalisations}, whereby partial thermalisation occurs within energy subspaces \cite{ptherm_perry}. There is also the prospect of addressing the same problem using a more sophisticated treatment of thermalisation dynamics, for example by accounting for non-Markovian effects.

On the other hand, the arguments in section IIA which rule out reversible quasistatic processes extend readily, and we expect that qualitatively similar results may hold for quantum operations of higher complexity. At a basic level, more input states will always mean a more constrained thermodynamic optimisation, and can only mean greater energy penalties: in this sense, much of the insight can already be gained from the two input case analysed here.

Sometimes, quantum investigations uncover analogous classical effects \cite{Smith}. Arguably the only truly quantum principle in play here is the inability to freely copy or measure the state\footnote{Necessary ingredients of Bennett's reversible computation \cite{bennett73} and of unitary quantum computation, if the result is to be read out.} and choose the protocol accordingly. We might imagine an analogous classical scenario where practical constraints prevent us from incorporating measurement feedback in our process: in that case we anticipate that a similar thermodynamic handicap might apply.

As we have seen, optimising a quantum thermodynamic process for more than one input is more costly than the free energy difference would suggest. Conversely, holding prior knowledge of the system's state confers a greater advantage, emphasising the fuel value of information \cite{delrio, Park}. This adds to a growing recognition that the Clausius inequality and Landauer bound do not tell the whole story with respect to irreversibility \cite{maroney,taranto,Violaris_2022,Kolchinsky,Dahlsten_2011}. 

\subsection*{Acknowledgements}
We would like to thank O. Maroney, P. Camati, J. Monsel, M. Violaris and F. Meng for stimulating discussions. This work was supported by the Engineering and Physical Sciences Research Council (EP/T518049/1).
FC and JT-B gratefully acknowledge funding from the Foundational Questions Institute Fund (FQXi-IAF19-01).
JA gratefully acknowledges funding from EPSRC (EP/R045577/1) and the Deutsche Forschungsgemeinschaft (DFG 384846402). JA
thanks the Royal Society for support.

\textcolor{\myblue}{\textit{For the purpose of open access, the authors have applied a ‘Creative Commons Attribution’ (CC BY) licence to any Author Accepted Manuscript version arising from this submission}}


%

\onecolumngrid
\appendix

\section{Deriving the bound on entropy production}\label{appx}
Applying the Cauchy-Schwarz inequality to \eqref{pinsker}, the entropy production $\Sigma$ can be bounded by:
\begin{equation}\label{cauchy}
    \Sigma \geq \frac{1}{2}\sqrt{\sum_{i=1}^N \frac{1}{(1-\lambda_i)^2}}\sqrt{\sum_{i=1}^N {\norm{\overline{\rho}_{i+1}-\overline{\rho}_i}_1}^4}.
\end{equation}
The first square-root term in the above can be bounded using the power mean inequality and the fact that $\ln x \leq x-1$ for all $x> 0$:
\begin{equation}\label{powermean}
    \begin{split}
        \left(\frac{1}{N}\sum_{i=1}^N (1-\lambda_i)^{-2}\right)^{-\frac{1}{2}} &\leq\frac{1}{N}\sum_{i=1}^N (1-\lambda_i)\\
        &=1-\frac{1}{N}\sum_{i=1}^{N}\lambda_i\\
        &\leq 1 - \prod_{i=1}^N \lambda_i^\frac{1}{N}\\
        &\leq -\frac{1}{N}\ln(\prod_{i=1}^N \lambda_i)\\
    \end{split}
\end{equation}
\begin{equation*}
    \implies \sqrt{\sum_{i=1}^N \frac{1}{(1-\lambda_i)^{2}}} \geq \frac{N\sqrt{N}}{-\ln(\prod_{i=1}^N \lambda_i)}.
\end{equation*}
Turning to the second square-root term in (\ref{cauchy}), we employ the fact that for any collection of positive reals $a_i$, $\sum_{i=1}^N a_i^2 \geq \frac{1}{N}\left(\sum_{i=1}^N a_i\right)^2$:
\begin{equation}\label{quartic}
    \begin{split}
        \sqrt{\sum_{i=1}^N {\norm{\overline{\rho}_{i+1}-\overline{\rho}_i}_1}^4} &\geq \frac{1}{\sqrt{N}}\sum_{i=1}^N {\norm{\overline{\rho}_{i+1}-\overline{\rho}_i}_1}^2\\
        &\geq \frac{1}{N\sqrt{N}}\left(\sum_{i=1}^N \norm{\overline{\rho}_{i+1}-\overline{\rho}_i}_1\right)^2\\
        &\geq \frac{1}{N\sqrt{N}} {\norm{\overline{\rho}_{N+1}-\overline{\rho}_1}_1}^2,
    \end{split}
\end{equation}
where the last line expresses the triangle inequality. Substituting (\ref{powermean}) and (\ref{quartic}) into (\ref{cauchy}):
\begin{equation}\label{almost}
    \Sigma \geq \frac{{\norm{\overline{\rho}_{N+1}-\overline{\rho}_1}_1}^2}{-2\ln(\prod_{i=1}^N \lambda_i)}.
\end{equation}
Finally, we note that $\overline{\rho}_1 = \overline{\rho}$ and that $U\,\overline{\rho}_{N+1}\, U^\dagger = \overline{\eta}$, where $U$ is given by equation \eqref{U}; and that $\prod_{i=1}^N \lambda_i = \frac{\norm{\eta_1-\eta_2}_1}{\norm{\rho_1-\rho_2}_1}$ \eqref{lambda}. This allows us to express \eqref{almost} in a way that depends only on the input and output states:
\begin{equation}\label{oldfinalbound}
    \Sigma \geq \frac{1}{2}\frac{{\norm{U^\dagger\overline{\eta} U - \overline{\rho}}_1}^2}{\ln(\frac{\norm{\rho_1-\rho_2}_1}{\norm{\eta_1-\eta_2}_1})},
\end{equation}
finally leading to an upper bound on extractable work:
\begin{equation}
\begin{split}
    \widetilde{W} 
    &\leq -\Delta \overline{F} - \frac{k_\mathrm{B} T}{2}\frac{{\norm{U^\dagger\overline{\eta} U - \overline{\rho}}_1}^2}{\ln(\frac{\norm{\rho_1-\rho_2}_1}{\norm{\eta_1-\eta_2}_1})}
\end{split}.
\end{equation}

\end{document}